\documentstyle[prl,aps,twocolumn,floats]{revtex}
\begin{document}
\draft
\title
{Towards the Final State of Spherical Gravitational Collapse and Likely
Source of Gamma Ray Bursts}
\author{Abhas Mitra}
\address{Theoretical Physics Division, Bhabha Atomic Research Center,\\
Mumbai-400085, India\\ E-mail: amitra@apsara.barc.ernet.in}


\maketitle

\begin{abstract}
By  analysizing the {\em global properties} of the equations for general
relativistic  collapse which reflect  the rate of change of the {\em
circumference radius} $R$ with the {\em proper radius} $l$, we show  that
(i) no ``trapped surface'' is ever formed  and (ii) the {\em gravitational
mass} of the fluid $M(r,t)\rightarrow 0$ as $R\rightarrow 0$. This result
is absolutely independent on the equation of state of the fluid or any
other details. Thus, for continued collapse, the total energy output
measured by a distant observer, is $Q\rightarrow M_ic^2$, the original gravitational
mass.

\end{abstract}

\pacs{PACS numbers: 04.20.Dw, 04.70. Bw}

The problem of gravitational collapse has been described as ``the greatest
crisis in physics of all time''\cite{1}(pp.1196).
 Though the theoretical framework for studying
this problem is fairly well developed\cite{2,3}, in practice it is
impossible to make much headway without making a number of simplifying
assumptions because of our inability to {\em self consistently} handle:
(i) the equation of state (EOS) of matter at arbitrarily high density and
temperature, (ii) the opacity of nuclear matter at such likely unknown
extreme conditions, (iii) the associated radiation transfer problem and
all other highly nonlinear and coupled partial differential equations. As
far as exact solution is concerned, it is possible only if the collapsing
fluid is considered to be a {\em dust} with pressure $p\equiv 0$
everywhere including the center even for {\em arbitrary high density}.
Further, it should be considered {\em homogeneous}, and the pioneering
solution to this effect was obtained by Oppenheimer and Snyder\cite{4}(OS)
which suggested that the spherical dust ball collapses to a point to a
state of infinite proper density, i.e., $R\rightarrow 0$, $\rho\rightarrow
\infty$ in a {\em proper time} ({\em by fiat, pressure still remains
zero}):
\begin{equation}
\tau={\pi \over 2} \left({3\over 8\pi \rho(0)}\right)^{1/2}
\end{equation}
where $\rho(0)$ is the initial mass-energy density of the fluid when it is
assumed to be {\em at rest}. A distant inertial observer $S_{\infty}$,
however can not witness this approach to a physical singularity because as
$R\rightarrow R_{\rm gi}= 2G M_i/c^2$, where $M_i$ is the initial
gravitational mass of the dust, the photons emitted from the suface of the
collapsing fluid are infinitely redshifted indicating the formation of an
``event horizon'' and a subsequent black hole (BH). Attempts for numerical
studies\cite{5} to study the collapse of a physical fluid can progress
only by presuming that the energy emitted during the process $Q\ll M_i
c^2$.  The said assumption tantamounts to the condition that the eventual
temperature of the nuclear matter $T\ll 1$GeV, the typical nucleon mass.
When this is so, one can use a $T=0$ degenerate EOS or can incorporate a
finite $T$ EOS which incorporates a first order correction term $\propto
T^2$. A general overall self-consistent numerical treatment of the problem
which does not impose explicit or tacit smallness conditions on $Q$ and
$T$ is implausible. For a dust, thermodynamics would tell that a $p=0$
condition implies that the internal energy density $e=0$ and also $T=0$.
Consequently, the instantaneous value of $Q\equiv 0$, which in turn
implies the instantaneous gravitational mass of the body $M\equiv M_i$.
The presumption that $Q\ll M_i c^2$ leads to a more or less similar
condition, that $M\approx M_i$, and naturally numerical computations
reveal the formation of an event horizon enclosing a mass $M_{\rm
g}=M_f\approx M_i$ within a finite proper time. However, when the
condition that for all times $Q\ll M_i c^2$ is not imposed the running
value of $M=M_i -Q/c^2< M_i$, consequently, the would be value of $R_{\rm
g}$ will be smaller than the initially suspected value of $R_{\rm gi}$
leading to value of $Q$ higher than the previous value. This may in turn
lead to a value of $M$ lower than its previously expected value. And, in general, it
can not be predicted whether this apparently runaway process converges to
a certain finite value of $M_{\rm g}$ in a finite time or not, where
$M_{\rm g}$ is the mass at $R=R_{\rm g}$. In this situation, we
would like to explore the limiting properties of the equations for
collapse for a perfect fluid {\em without making any conceptual assumption}.
 The
matter part of the fluid has the energy momentum tensor
\begin{equation}
T^{\mu \nu} = (p+\rho) u^\mu u^\nu +p g^{\mu \nu}
\end{equation}
where $u^\mu$ is the fluid 4-velocity, and $\rho=\rho_0+e$, with $\rho_0=m
n$,  $m$ is the rest mass of a baryon as measured by $S_{\infty}$ and $n$
is the proper number density of the baryons. Now we have taken $c=1$. The
radiation part of the fluid is described by
\begin{equation}
E^{\mu \nu}= q k^\mu k^\nu
\end{equation}
where $q$ is both the energy density and the radiation flux in the proper
frame and $k^\mu =(1;1,0,0)$ is a null geodesic vector\cite{6} so that $k^\mu k_\mu=0$.
Irrespective of whether a ``trapped surface'' is formed or not,
the interior of the collapsing body may be described by the Schwarzschild
metric\cite{1,2,3,6}:
\begin{equation}
ds^2=-e^{2\phi} dt^2 +e^{2\lambda} dr^2 + R(r,t)^2
(d\vartheta^2+\sin^2\vartheta d\varphi^2)
\end{equation}
Here $r$ is a comoving coordinate, so that the total number of baryons
$N(r)$ enclosed within the sphere of constant coordinate radius $r$ is
constant and independent of $t$ and $R(r,t)$
is the circumference coordinate. The explicit forms of $e^\phi$ and
$e^\lambda$ are not known, but they are nonsingular for any finite $R$, and
computational techinques are supposed to self-consistently evaluate them.
 The exterior region must be described in
terms of Vaidya's retarded time $u$\cite{7}, which turns out to be the proper time
of $S_\infty$\cite{3}:
\begin{equation}
ds^2=- e^{2\psi}  du^2 -2 e^\psi e^\lambda du~dr +R^2(d\vartheta^2+ \sin^2
\vartheta d\varphi^2)
\end{equation}
where $e^{2\psi}=(1-2GM(u)/ R)$.
The comoving  time and radial derivative operations are respectively
defined by
\begin{equation}
D_{\rm t}=e^{-\phi}\left({\partial \over
\partial t}\right)_{\rm r}={d\over d\tau}; \qquad D_{\rm
r}=e^{-\lambda}\left({\partial\over \partial r}\right)_{\rm t}={d\over dl}
\end{equation}
Here $d\tau=e^\phi dt$
is an element of proper time at constant $r$, and
$dl=e^{\lambda} dr$ is an element of proper radial distance between two
baryon-shells $r$ and $r+dr$, $\vartheta$ and
$\varphi$ are frozen in spherical symmetry.
The {\em circumference velocity} of the fluid measured at
constant $r$ in terms of proper time\cite{2,3} is
\begin{equation}
D_{\rm t}R=U ={dR\over d\tau}
\end{equation}
Here $D_t$ and $D_r$ are to be treated as equivalents of ``convective derivative''
following the trajectory of the fluid parcel used in classical hydrodynamics.
 The central {\em global
character} of the collapse process, {\em valid in both metrics}, is manifest through
\begin{equation}
\Gamma=D_{\rm r}R={d(circumference)/2\pi\over d(proper~
distance)}={dR\over dl}
\end{equation}
And the most important {\em global constraint} on the field equations is
that the relationship
\begin{equation}
\Gamma^2=1- {2GM\over R} +U^2
\end{equation}
is {\em preserved for all times}\cite{2,3}.
The actual equations for containing the details of the collapse process,
and {\em valid in metric} (4),  are \cite{3}
\begin{equation}
D_{\rm r}M=4\pi R^2\left[\Gamma (\rho +q)+ U q\right]
\end{equation}
\begin{equation}
D_{\rm t}M =- 4 \pi R^2 p U- L (U+\Gamma)
\end{equation}
\begin{equation}
D_{\rm t}U=-{\Gamma \over \rho+p} \left({\partial p \over \partial
R}\right)_{\rm t} -{M+4\pi R^3 (p+q)\over R^2}
\end{equation}
\begin{equation}
D_{\rm t}\Gamma=-{U\Gamma\over \rho +p}\left({\partial p\over \partial
R}\right)_{\rm t} +{L\over R}
\end{equation}
where the comoving luminosity is
\begin{equation}
L=4\pi R^2 q
\end{equation}
For motionless matter in flat space ($S_{\infty}$), $\Gamma\equiv 1$.
In a curved space $\Gamma<1$ because of the
gravitational field. And a value of $\Gamma<0$ {\em would signify that the
spacetime is pinched off from the rest of the universe}\cite{3}, and this
is what is believed to happen in the final stages of gravitational
collapse. However {\em nobody has ever hinted what could be order of
magnitude of the (-ve) value of final $\Gamma_f$ even for the idealized
case of a dust collapse}. The popular conviction that even a physical
fluid must eventually behave like a O-S dust is based on certain
properties of the {\em radiationless} collapse equations\cite{2}:
\begin{equation}
D_{\rm r}M=4\pi R^2 \rho \Gamma
\end{equation}
\begin{equation}
D_{\rm t}M= -4\pi R^2 p U
\end{equation}
\begin{equation}
D_{\rm t} U=-{\Gamma\over \rho+p} \left({\partial p\over \partial R
}\right)_{\rm t} -{M+4 \pi R^3 p \over R^2}
\end{equation}
\begin{equation}
D_{\rm t} \Gamma=-{U \Gamma \over \rho +p}\left({\partial p \over \partial
R}\right)_{\rm t}
\end{equation}
The traditional argument favouring BH formation goes like this\cite{8}:
For collapse, $U<0$, and from eq. (16) it follows that $M(r,t)$ increases
monotonically within the fluid; of course, it remains constant, for
adiabatic collapse, for $R\ge R_0$, the outer boundary. Because of this,
the $M/R^2$ term in eq. (17) increases more rapidly than than it would in
the Newtonian theory where it remains constant. Then eq.(18) would
suggest that $\Gamma$ decreases monotonically instead of remaining
constant in a Newtonian case. Both of these relativistic effects are
supposed to cause the collapse process monotonic and accelerate faster
than the Newtonian case\cite{8}.  Here note that unless we presume that
$\partial p/\partial R$ can be positive,  a negative value of $\Gamma$ in
eq.(18) would demand that $\Gamma$ increases with time implying $\Gamma$
{\em would tend to be back to zero}. In any case when we restore $q$, or in other
words, when we do away with the assumption $Q\ll M_i $, {\em this entire line
of argument collapses}  and nothing can be predicted about the nature of
evolution of the collapse process.  In this situation, we discuss below
the global properties  of these equations. Following the
discussion by Landau \& Lifshitz\cite{9}(pp.250), the {\em ordinary radial velocity of the
fluid as measured in a local rest frame} is
\begin{equation}
v^{\hat r}\equiv v={e^\lambda\over e^\phi} {d r\over d t}\equiv {dl\over d\tau}
\end{equation}
By {\bf principle of equivalence} (POE), $v\le c$. By combining eq. (7),
 (8), and (20), we find
\begin{equation}
U=\Gamma v
\end{equation}
The above relationship does not depend on the signature or form of the
metric coefficients and, as mentioned earlier, all equations involving {\em interior
Sch. metric coefficients} remain vaild even if a trapped surface is formed\cite{6}.
Further, eq. (20) {\em must not be confused with} $u^{\hat r}=\gamma v$, where
$u^{\hat r}$ is the radial component of the 4-velocity measured in the
local inertial frame (in the comoving Lagrangian frame, $u^r\equiv 0$) and
$\gamma=(1-v^2)^{-1/2}$ is the Lorentz Factor of the fluid in the same
inertial frame. By combining eq. (9) and (19), we obtain
\begin{equation}
{\Gamma^2\over \gamma^2}\equiv 1-{2GM\over R}
\end{equation}
Since the LHS of the foregoing eq. is positive (even if $\Gamma$ were
negative), so will be the RHS, and therefore, {\bf by simply using POE},
we find that
\begin{equation}
{2GM\over R} \le 1
\end{equation}
Recall that, for spherical collapse the condition for formation of a {\em
trapped surface}, a surface of no-return is $2GM/R >1$\cite{3,10}. Thus, in a most
general manner, we find that {\em trapped surfaces are not allowed by GTR}
contrary to the assumtion made in a large number of previous works
staring from the sixties. It is the assumption (1) about the occurrence of the
 trapped surfaces, alongwith, (2) the causality conditions requiring the
nonexistence of closed timelike lines, (3) the strong energy condition
 $(T_{\mu\nu} -0.5 g_{\mu\nu})u^{\mu} u^{\nu}\ge 0$, and (4) a certain generality condition
on the Riemann-Christoffel tensor, that  laid the  foundation of the
famous singularity theorems of Hawking, Penrose, Geroch and others\cite{10}.
 So now they are invalidated because trapped surfaces are not allowed by GTR.
 We quickly recall the relevant comments by
Misner et al.\cite{1}(pp.935): ``All the conditions except the trapped surface seem
eminently reasonable for any physically realistic space time''.  Thus,
the  mathematical and moral foundation for the inevitability of
occurrence of singularities in GTR are eliminated. However, this does not
necessarily mean that singularities must not occur, and, we will probe this
question in an accompanying paper\cite{10}. Any EOS will
support only a certain finite maximum $M_i$ and therefore given sufficiently large $M_i$,
continued collapse is a genuine possibility; and if
indeed $R\rightarrow 0$, then  in order to
satisfy the foregoing ineqality, {\em we must have}
\begin{equation}
M_f\rightarrow 0; \qquad R_f\rightarrow 0
\end{equation}
Remember here that the quantity  $M_0 =m N$ (which is the baryonic mass of
the star, if there are no antibaryons) is conserved as $M_f \rightarrow 0$.
How do recoincile our general result $M_f=0$,
 with the result of O-S collapse which suggests that
as the dust ball with an initial condition $R=R_0$, $M(r_0, 0)=M_i$
collapses from a {\em state of rest}, i.e., $v=v(r,0)=0$ and $U=U(r,0)=0$,
it forms a BH with $M_f=M_i$ in a finite proper time? To find the
 $\rho(0)$ occuring in eq. (1) in a truly self-consistent way, we must fix
it from the collapse equations themselves, rather than {\em artificially and
by hand}.
When we put
this $U=q=0$ condition in eq.(12), we obtain the so-called
Oppenheimer-Volkov (OV) equation\cite{11}.
\begin{equation}
{dp\over dR}=- {p+\rho(0) \over R(R-2GM)}(4\pi p R^3 + 2G M)
\end{equation}
By feeding the $p=0$ EOS in the OV
equation  it may be verified that
whether the fluid is considered homogeneous or not the solutions
could be of two types: (i)$ \rho(0)=0, ~R_0=\infty, ~M_i=finite$ and (ii)
$\rho(0)=0,~ R_0=finite, ~M_i=0$. In the former case, the fluid boundary
requires infinite $\tau$ for attaining  a finite $R$ or $R_g$, implying
that the collapse process hardly ever takes off:
\begin{equation}
\tau(R)={2\over3}\left[{R_0\over c}\left(R_0\over R_{\rm g}\right)^{1/2}-
{R\over c}\left(R\over R_{\rm g}\right)^{1/2}\right]
\end{equation}
and there is no question of formation of a BH.
And  in the latter case, our
proof that $M_f\equiv 0$ is vindicated because $M_f=M_i=0$.
Even in this case, eq. (1) yields $\tau
=\infty$! These {\bf boundary
conditions}  {\em are fudged in all
theoretical and numerical discussions by pretending that a self-supporting
fluid with finite values of $\rho$ and $R_0$, and $p\neq 0$, everywhere,
may approximately be considered to suddenly attain the dust-like
dynamics}.  Even if such a dust-like collapse took place, it can be
verified from eq.(18) (by evolving  from $U=0$ condition) that, far
from assuming any negative value,  the value of $\Gamma$ remains fixed to
the initial value $\Gamma_0$, which is unity, and which is the appropriate value of
$\Gamma$ for a motionless (no kinetic energy) and matter free (thus
gravityfree) region. It might be thought that if the numerical
calculations study the collapse of a supermassive star, the horizon may be
formed at very small densities $\sim 10^{-4}$g/cm$^3$, and the entire
collapse physics including the EOS can be handled accurately. This idea
would not be correct because, the expectation that horizon would be formed
at $R_{\rm g}\approx 2GM_i$ is based on the presumption that $Q\ll M_i$.
And in any case our central result manifest by eq. (22) is an exact one.
In the context of a numerical neutron star collapse, we will
discuss about an ansatz for handling the thremodynamics of the problem
which may reveal a value of $Q$ much higher than previously reported
values, although the eventual computation must break down as $T_{\rm com}$
keeps on increasing without any upper limit\cite{13}.

  It is widely believed that Chandrasekhar's
discovery that White Dwarfs (WD) can have a maximum mass set the stage for
having a gravitational singular state with finite mass. The hydrostatic
equilibrium of WDs can be approximately described by Newtonian
polytropes\cite{14} for which one has $R\propto \rho_c^{(1-n)/2n}$, where
$\rho_c$ is the central density of the polytrope having an index $n$.  It
shows that, for a singular state i.e., for $\rho_c\rightarrow \infty$, one
must have $R\rightarrow 0$ for $n>1$; and Chandrasekhar's limiting WD
indeed has a {\em zero radius}. On the other hand, the mass of the
configuration $M\propto \rho_c^{(3-n)/2n}$.  And unless $n=3$,
$M\rightarrow \infty$ for the singular state. However, for
ultrarelativistic electrons, the EOS is $p\rightarrow e/3$ and the
corresponding $n\rightarrow 3$ and one obtains a finite value of $M_{ch}$
- the Chandrasekhar mass.  Now when we apply the theory of polytropes for
a case where the pressure is supplied by the baryons and not only by
electrons, we must consider GTR polytropes\cite{15}. It can be easily
verified from eq. (2.24) of ref.(15) that in the limit $\rho_c
\rightarrow \infty$, the scale size of GTR polytropes $A^{-1} \rightarrow 0$.
Further eq.(2.15) and (4.7) of ref. (15)  tell that $M \propto K^{n/2}
\propto \rho_c^{-1/2} \rightarrow 0$ for $\rho_c\rightarrow \infty$. Thus,
a {\em proper GTR extension of Chandrasekhar's work would not lead to a BH
of finite mass}, but, on the other hand, to a singular state with
$M\rightarrow 0$.  Physically, the $M=0$ state may result when the
{\em negative gravitational energy} exactly cancels the internal energy, the
{\em baryonic mass energy} $M_0$ and any other energy, and  which
is possible in the limit $\rho \rightarrow \infty$ and $p\rightarrow
\infty$.  In fact, Zeldovich and Novikov\cite{12} discussed the
possibility of having an ultracompact configuration of degenerate fermions
obeying the EOS $p=e/3$ with $M\rightarrow 0$; we have verified that their
configuration actually corresponds to $R\rightarrow R_{\rm g} \rightarrow
0$. In fact, they mentioned  the possibility of {\em having a machine for
which} $Q\rightarrow M_i c^2$.
In a different context, it has been argued that {\em naked singularities}
produced in spherical collapse must have $M_f=0$\cite{16}. Also, it was
discussed long ago\cite{15} that spherical gravitational collapse should
come to a {\em decisive end} with $M_f=M^*=0$, and, in fact, this
understanding was formulated as a ``Theorem''\cite{17}:

``THEOREM 23. Provided that matter does not undergo collapse at the
microscopic level at any stage of compression, then, -regardless of all
features of the equation of state - there exists for each fixed number of
baryons A a ``gravitationally collapsed configuration'', in which the
mass-energy $M^*$ as sensed externally is zero.'' We have proved this and
more for  spherical symmetry, our result is absolutely general and has got
nothing to with the exact EOS and radiation transport equations. On the
other hand, it depends only on the global properties of the GTR collapse
equations and the Principle of Equivalence. We have learnt that GRB990123
released an energy $3\times 10^{54}$ergs of energy in $\gamma$-rays alone
 under
condition of isotropy.
Given the fact that the relativistic blast wave
radiating the gamma-rays may not have an efficiency greater than 10-20\%
in converting the blast wave energy into gamma rays, the total
electromagnetic energy released in the process could easily be $Q_{FB}\sim
10^{55}$ erg.
Further, the energy output in $\nu$s could be or rather should be
$>Q_{FB}$, the net energy released in the process could be $Q\sim
few~10^{55}$ erg. In case, there is weak beaming, the total energy
released may still be $Q \le 10^{55}$ erg.
physical mechanism, other than what has been propsed here, can
satisfactorily explain the origin of such unprecedented energy. Such
energy release is hardly possible if trapped surfaces would form.

Acknowledgement: I am thankful to Prof. P.C. Vaidya for painstakingly
going through this work, offering some suggestions, and confirming that
this work is physically and general relativistically correct.

\end{document}